\documentclass[preprint,prd,aps,tighten,nofootinbib,amssymb]{revtex4}
\usepackage{epsf,epsfig,graphicx,subfigure}
\usepackage{amssymb,amsmath}
\usepackage{slashed, color}

\newcommand{\beq}{\begin{equation}}
\newcommand{\eeq}{\end{equation}}
\newcommand{\bea}{\begin{eqnarray}}
\newcommand{\eea}{\end{eqnarray}}
\newcommand{\bear}{\begin{array}}
\newcommand {\eear}{\end{array}}
\newcommand{\bef}{\begin{figure}}
\newcommand {\eef}{\end{figure}}
\newcommand{\bec}{\begin{center}}
\newcommand {\eec}{\end{center}}

\newcommand{\dis}[1]{\begin{equation}\begin{split}#1\end{split}\end{equation}}

\begin{document}
\draft
\tighten
\preprint{CTPU-16-13}

\title{\large \bf 750 GeV diphoton resonance  and electric dipole moments}

\author{
    Kiwoon Choi$^{a}$\footnote{email: kchoi@ibs.re.kr},
    Sang Hui Im$^{a}$\footnote{email: shim@ibs.re.kr},
     Hyungjin Kim$^{a, b}$\footnote{email: hjkim06@kaist.ac.kr},
     Doh Young Mo$^{a}$\footnote{email: modohyoung@ibs.re.kr}}
     
\affiliation{
 $^a$Center for Theoretical Physics of the Universe, \\
 Institute for Basic Science (IBS), Daejeon 34051, Korea \\ 
 $^b$Department of Physics, KAIST, Daejeon 34141, Korea 
    }

\vspace{4cm}

\begin{abstract}
We examine the implication of the recently observed 750 GeV diphoton excess for the electric dipole moments of the neutron and electron. If the excess is due to a spin zero resonance which couples to photons and gluons through the loops of massive vector-like fermions, the resulting neutron  electric dipole moment can be comparable to the present experimental bound if the CP-violating angle $\alpha$ in the underlying new physics  is of ${\cal O}(10^{-1})$. An electron EDM comparable to the present bound can be achieved through a mixing between the 750 GeV resonance and the Standard Model Higgs boson,  if the mixing angle itself for an approximately pseudoscalar resonance, or the mixing angle times the CP-violating angle $\alpha$ for an approximately scalar resonance, is of ${\cal O}(10^{-3})$. For the case that the 750 GeV resonance  corresponds to a composite pseudo-Nambu-Goldstone boson formed by  a QCD-like hypercolor dynamics confining at $\Lambda_{\rm HC}$, the resulting neutron EDM can be estimated with   $\alpha\sim (750 \, {\rm GeV}/\Lambda_{\rm HC})^2\theta_{\rm HC}$, where $\theta_{\rm HC}$ is the hypercolor vacuum angle.
\end{abstract}

\maketitle

\section{Introduction}

Recently
the ATLAS and CMS collaborations reported an excess of diphoton events at the invariant mass $m_{\gamma\gamma} \simeq  750$ GeV with the local significance 3.6 $\sigma$ and 2.6 $\sigma$, respectively \cite{ATLAS, CMS}. The analysis was updated later, yielding an increased local significance, 3.9 $\sigma$ and 3.4 $\sigma$, respectively \cite{ATLAS1, CMS1}. If the signal persists, this will be an unforeseen discovery of new physics beyond the Standard Model (SM).  So one can ask now what would be the possible phenomenology other than the diphoton excess,  which may result from the new physics  to explain the 750 GeV diphoton excess.

With the presently available data,  one simple scenario  to explain the diphoton excess is a SM-singlet spin zero resonance $S$  which couples to massive vector-like fermions carrying non-zero SM gauge charges \cite{Franceschini:2015kwy}. In this  scenario, the 750 GeV resonance interacts with the SM sector dominantly  through the SM gauge fields and possibly also through the Higgs boson. In such case, if the new physics sector involves a CP-violating interaction,   the electric dipole moment (EDM) of the neutron or electron may provide the most sensitive probe of new physics in the low energy limit.

More explicitly, after integrating out the massive vector-like fermions, the effective lagrangian  may include  
\dis{ \label{1PI_general}
\frac{\kappa_s}{2}SF^{a\mu\nu}F^a_{\mu\nu}+\frac{\kappa_p}{2} SF^{a\mu\nu}\tilde F^a_{\mu\nu}
+\frac{d_W}{3} f_{abc} F^a_{\mu \rho} F_\nu^{b\,\,\rho} \widetilde{F}^{c \mu \nu}
+ ...,
}
where $F^a_{\mu\nu}$  denotes the SM gauge field strength and $\tilde F^a_{\mu\nu}=\frac{1}{2}\epsilon_{\mu\nu\rho\sigma}F^{a\mu\nu}$ is its dual. In view of that the SM weak interactions break CP explicitly through the complex Yukawa couplings\footnote{Throughout this paper, we assume the CP invariance in the strong interaction is due to the QCD axion associated with a Peccei-Quinn $U(1)$ symmetry \cite{Kim:2008hd}}, it is quite plausible that the underlying dynamics of $S$ generically breaks CP, which would  result in nonzero value of the effective couplings $\kappa_s\kappa_p /\sqrt{\kappa_s^2+\kappa_p^2}$ and $d_W$. As is well known, in the presence of those CP violating couplings,  a nonzero neutron or electron EDM can be induced through the loops involving the SM gauge fields \cite{Weinberg:1989dx,Barr:1990vd,Marciano:1986eh,Boudjema:1990dv}.

In this paper, we examine the neutron and electron EDM in models for the 750 GeV resonance, in which the effective interactions  (\ref{1PI_general}) are generated by the loops of massive  vector-like fermions. We find that for the parameter region to give the diphoton cross section $\sigma (pp\rightarrow \gamma\gamma)= 1\sim 10$ fb, the neutron EDM can be comparable to the present experimental bound,  e.g. $d_n \sim {\rm a\,\,few}\times 10^{-26}\,\, {\rm e\cdot cm}$, if the CP-violating angle $\alpha$ in the underlying dynamics is of  ${\cal O}(10^{-1})$, where $\sin 2\alpha \sim \kappa_s\kappa_p /\sqrt{\kappa^2_s+\kappa_p^2}$ in terms of the effective couplings in (\ref{1PI_general}).
An electron EDM near the present bound can be obtained also 
through a mixing between $S$ and the SM Higgs boson $H$.
We find that again  for the parameter region of $\sigma (pp\rightarrow \gamma\gamma)= 1\sim 10$ fb, the electron EDM is given by 
 $d_e\sim  6\times 10^{-26} \sin\xi_{SH}\sin\alpha\,\, {\rm e\cdot cm}$, where
$\xi_{SH}$ is the $S-H$ mixing angle\footnote{Note that if $\xi_{SH}$ corresponds to a CP-violating mixing angle, then  $\sin\alpha$ in this expression  is not a CP-violating parameter anymore, and therefore is a parameter of order unity.}. 
Our result on the neutron EDM can be applied also to the models in which $S$ corresponds to a composite pseudo-Nambu-Goldstone boson formed by  a QCD-like hypercolor dyanmics which is confining at $\Lambda_{\rm HC}$ \cite{Harigaya:2015ezk,Nakai:2015ptz,Redi:2016kip}. In this case, the CP-violating order parameter
$\alpha$ can be identified as $\alpha\sim \theta_{\rm HC}m_S^2/\Lambda^2_{\rm HC}$, where $\theta_{\rm HC}$ denotes the vacuum angle of the underlying QCD-like hypercolor dynamics.

The organization  of this paper is as follows. In section \ref{sec:model}, we introduce a simple model for the 750 GeV resonance involving CP violating interactions, and summarize the diphoton signal rate given by the model. 
In section \ref{sec:EDM}, we examine  the neutron and electron EDM  in the model of section \ref{sec:model}, and discuss the connection between the resulting EDMs and the diphoton signal rate. Although we are focusing on a specific model, our results can
be used for an estimation  of EDMs in more generic models for the 750 GeV resonance. 
In section \ref{sec:composite},  we apply our result to the case that $S$ is a composite pseudo-Nambu-Goldstone boson formed by a QCD-like hypercolor dynamics.   
Section \ref{sec:conc} is the conclusion.

\section{A model for diphoton excess with CP violation} \label{sec:model}

The 750 GeV diphoton excess can be explained most straightforwardly by introducing a SM-singlet spin zero resonance $S$ which couples to massive vector-like fermions to generate the effective interactions (\ref{1PI_general}) \cite{Franceschini:2015kwy}.
To be specific, here we consider a simple model
involving $N_F$  Dirac fermions $\Psi=(\Psi_1, \Psi_2, ..., \Psi_{N_F})$ carrying a common charge under the SM gauge group $SU(3)_c\times SU(2)_L\times U(1)_Y$.
 Then the most general renormalizable interactions of $S$ and $\Psi$ include
\bea
\label{toy model}
{\cal L} = \bar{\Psi} i \slashed{D} \Psi -  \bar{\Psi} \left( M + Y_s S   +i Y_p S \gamma^5 
\right)\Psi-\frac{1}{2}m_S^2 S^2 -A_{SH}S|H|^2 + ...,
\eea
where the mass matrix $M$ can be chosen to be real and diagonal, while  $Y_{s,p}$ are hermitian Yukawa coupling matrices. Here $H$ is the SM Higgs doublet, and we have chosen the field basis for which $S$ has a vanishing vacuum expectation value in the limit to ignore its mixing with $H$.
For simplicity, in the following we assume that all fermion masses and the Yukawa couplings are approximately flavor-universal, so they can be parametrized as 
\bea
M \,\approx \, m_\Psi \, {\bf 1}_{N_F\times N_F}, \quad
Y_s\approx y_S \cos\alpha\,{\bf 1}_{N_F\times N_F} , \quad Y_p \approx y_S \sin\alpha\, {\bf 1}_{N_F\times N_F}, \eea
where ${\bf 1}_{N_F\times N_F}$ denotes the $N_F\times N_F$ unit matrix.
Note that in this parametrization  $\sin 2\alpha$ corresponds to the order parameter for CP violation. In the following, we will often use $\alpha$ (or $\sin\alpha$) as a CP violating order parameter, although it should be $\alpha -\pi/2$ (or $\cos\alpha$) for an approximately  pseudoscalar $S$.

Under the above assumption on the model parameters, one can compute the 1PI amplitudes for the production and decay of $S$ at the LHC, yielding \cite{Franceschini:2015kwy}
\dis{ \label{effective}
{\cal L}_{\rm 1 PI} &=   
\frac{g_3^2}{16\pi^2 m_S} S \left( c^{(s)}_3 G^a_{\mu\nu} G^{a\mu\nu}  + c^{(p)}_3 G^a_{\mu\nu} \widetilde{G}^{a\mu\nu} \right)
 \\
&+ \frac{g_2^2}{16\pi^2 m_S} S \left( c^{(s)}_2 W^a_{\mu\nu} W^{a\mu\nu} + c^{(p)}_2 W^a_{\mu\nu} \widetilde{W}^{a\mu\nu} \right)
\\ 
&+ \frac{g_1^2}{16\pi^2 m_S} S \left( c^{(s)}_1 B_{\mu\nu} B^{\mu\nu} + c^{(p)}_1 B_{\mu\nu} \widetilde{B}^{\mu\nu} \right),
}
where
\dis{ \label{1PI_coupl}
c_i^{(s)} &= N_F y_S \cos \alpha \,  {\rm Tr}(T_i^2(\Psi))  \frac{m_S}{m_\Psi} \frac{A_{1/2}(\tau_\Psi)}{2}, \\
c_i^{(p)} &= -N_Fy_S \sin \alpha \,    {\rm Tr}(T_i^2(\Psi))  \frac{m_S}{m_\Psi} \frac{f(\tau_\Psi)}{\tau_\Psi},
}
with $i= 1,\, 2,\, 3$ denoting the SM gauge groups $ U(1)_Y,\, SU(2)_L,\, SU(3)_c$, respectively, 
and $\tau_\Psi \equiv m_S^2/4m_\Psi^2$. 
The loop functions $A_{1/2}(\tau)$ and  $f(\tau)$ are given by 
\bea
A_{1/2}(\tau)& =& 2 \big[ \tau + (\tau -1 )f(\tau)]/\tau^2,
\nonumber \\
f(\tau) &=& -\frac{1}{2} \int_0^1 dx \frac{1}{x} \ln [1 -4 x(1-x) \tau] \nonumber \\
&=& 
\left\{ \begin{array}{ll}
(\arcsin \sqrt{\tau})^2, & \tau \leq 1 \\
-\frac{1}{4}\left[ \ln \left( \frac{1+ \sqrt{1-\tau^{-1}} }{1- \sqrt{1-\tau^{-1}} } \right) - i \pi \right]^2, & \tau>1.
\end{array}
\right.\label{A-f}
\eea
Note that with a nonzero value of the CP violating angle $\alpha$,  the 750 GeV resonance $S$ couples to both $F^a_{\mu \nu} F^{a \mu \nu}$ and $F^a_{ \mu \nu} \widetilde{F}^{a \mu \nu}$. These two couplings turn out to incoherently contribute to the decay rate of $S$, so that the relevant decay rates are given by 
\bea \label{decay_phph}
\Gamma_{\gamma\gamma} &=& \frac{1}{4\pi} \left(\frac{e^2}{16\pi^2}\right)^2 m_S \left(\left| c_\gamma^{(s)}\right|^2 +\left|c_\gamma^{(p)} \right|^2 \right),
\\
\Gamma_{gg} &=& \frac{8}{4\pi} \left(\frac{g_3^2}{16\pi^2}\right)^2 m_S \left(\left| c_g^{(s)}\right|^2 +\left|c_g^{(p)} \right|^2 \right), \label{decay_gg}
\eea
in the rest frame of $S$. The diphoton signal cross section at the LHC can be estimated using the narrow width approximation \cite{Franceschini:2015kwy}, yielding
\bea 
\sigma(pp\rightarrow S \rightarrow \gamma\gamma) = C_{gg} \frac{1}{s} \frac{m_S}{\Gamma_S} \frac{\Gamma_{\gamma\gamma}}{m_S} \frac{\Gamma_{gg}}{m_S},
\eea
where the coefficient $C_{gg} = 2137$ at $\sqrt{s} = 13$ TeV, and $\Gamma_S$ denotes the total decay width of $S$.
Manipulating this, the decay rate should satisfy the following relation,
\dis{ \label{sig_reg}
\frac{\Gamma_{\gamma\gamma}}{m_S} \frac{\Gamma_{gg}}{m_S} 
=  2.17 \times 10^{-9} \left( \frac{\Gamma_S}{1 \, {\rm GeV}} \right) \left( \frac{\sigma_{\rm signal}}{8 \,{\rm fb}} \right),
}
for the signal cross section $\sigma = 1 \sim 10$ fb.
Here we normalize the total decay rate of $S$ by $\Gamma_S = 1$ GeV, since it is a typical value when there is an appreciable  mixing between the singlet scalar $S$
and the SM Higgs doublet \cite{Falkowski:2015swt}.

Plugging (\ref{1PI_coupl}) and (\ref{decay_phph}, \ref{decay_gg}) into  (\ref{sig_reg}), we obtain a relation
which is useful for an estimation of the electric dipole moments over the diphoton signal region:
\dis{ \label{m/y}
\frac{m_\Psi}{y_S} = 96\, {\rm GeV} \times Q_\Psi N_F \left(\frac{2}{3} {\rm Tr}(T_3^2(\Psi)) {\rm Tr}({\bf 1}(\Psi))\right)^{1/2}  \left( \frac{1\, \rm GeV}{\Gamma_S} \right)^{1/4} \left( \frac{8 \, \rm fb}{\sigma_{\rm signal}} \right)^{1/4} R_\Psi,}
where
$$ 
R_\Psi (\alpha, \tau_\Psi=m_S^2/4m^2_\Psi) = \left(\frac{c_\alpha^2 (A_{1/2}(\tau_\Psi)/2)^2 + s_\alpha^2 (f(\tau_\Psi)/\tau_\Psi)^2}{c_{0.1}^2 (A_{1/2}(1/4)/2)^2 + s_{0.1}^2 (4f(1/4))^2} \right)^{1/2} = \, {\cal O}(1). 
$$
Here $Q_\Psi$ and $T_3(\Psi)$ denote the electromagnetic and color charge of $\Psi$, respectively, ${\rm Tr}({\bf 1}(\Psi))$
is the dimension of the gauge group representation of $\Psi$, and $s_\alpha=\sin\alpha$ and $c_\alpha=\cos\alpha$. Note that $R_\Psi$ represents the dependence on $\tau_\Psi=m_S^2/4m^2_\Psi$ and $\alpha$, which is normalized to the value at
$\tau_\Psi = 1/4$ and $\alpha = 0.1$. As $R_\Psi$  has a mild dependence on $\tau_\Psi$ and $\alpha$,
the range of the parameter ratio $m_\Psi/y_S$ which would explain the diphoton excess can be easily read off from the above relation.

To see the origin of the CP violating angle $\alpha$, one may consider a UV completion of the model (\ref{toy model}). In regard to this, an attractive possibility is that  the model is embedded at some higher scales into a supersymmetric model including a singlet superfield $\phi$ and $N_F$ flavors of  vector-like charged matter superfields $\psi+\psi^c$~\cite{Hall:2015xds,Nilles:2016bjl}. The most general renormalizable superpotential of $\phi$ and $\psi+\psi^c$ is given by
\bea
\label{susy_model}
W =  (M + Y\phi)\psi\psi^c + \frac{1}{2}\mu_\phi \phi^2   + \frac{1}{3}\kappa \phi^3,
\eea
where without loss of generality $M$ can be chosen to be real and diagonal, ${\rm det}(Y)$ to be real, and  $\phi$ to have a vanishing vacuum value in the limit to ignore the mixing with the Higgs doublets.  Again, for simplicity let us assume that the mass matrix $M$ and the Yukawa coupling matrix $Y$ are approximately flavor-universal, and therefore
\bea
M \approx m_\Psi {\bf 1}_{N_F\times N_F}, \quad Y \approx y_S {\bf 1}_{N_F\times N_F}.\eea
Including the soft  supersymmetry (SUSY) breaking terms, the scalar mass term of $\phi$ is given by 
\bea
\left(|\mu_\phi|^2 + m_\phi^2\right) |\phi|^2 + \frac{1}{2}\left( B_\phi\mu_\phi \phi^2 + {\rm h.c}\right),
\eea
where $m_\phi$ is a SUSY breaking soft scalar mass, while $B$ is a holomorphic bilinear soft parameter. Note that in our prescription, both $\mu_\phi$ and $B_\phi$ are complex in general. 

Without relying on any fine tuning other than the minimal one to keep the SM Higgs to be light, one can arrange the SUSY model parameters to identify 
the lighter mass eigenstate of $\phi$ as the 750 GeV resonance $S$, and the fermion components of $\psi+\psi^c$ as the Dirac fermion $\Psi$ to generate the effective interactions (\ref{effective}), while keeping all other SUSY particles heavy enough 
to be in multi-TeV scales.
Then our model (\ref{toy model}) arises as a low energy effective theory at scales around TeV from the SUSY model (\ref{susy_model}),  with the matching condition
$$
\frac{1}{\sqrt{2}}S  = {\rm Re}(\phi)\cos\alpha + {\rm Im}(\phi)\sin\alpha,
$$
where 
\bea
\tan2\alpha = \frac{{\rm Im}(B_\phi\mu_\phi)}{{\rm Re}(B_\phi\mu_\phi)}.
\eea

Another possibility, which is completely different but equally interesting, would be that $S$ corresponds to a pseudo-Nambu-Goldstone boson formed by a QCD-like hypercolor dynamics which confines at scales near TeV. As we will see in section \ref{sec:composite},  the CP violating order parameter $\alpha$ in such models can be identified as
\bea
\sin 2 \alpha \sim \frac{m_S^2}{\Lambda_{\rm HC}^2}\sin \theta_{\rm HC},
\eea
where $\Lambda_{\rm HC}$ is the scale of spontaneous chiral symmetry breaking by the hypercolor dynamics and $\theta_{\rm HC}$ is the hypercolor vacuum angle.

\section{Electric Dipole Moments} \label{sec:EDM}

In this section, we estimate the electric dipole moments (EDMs) induced by the 750 GeV  sector in terms of the 
model introduced in the previous section.
At energy scales below $m_\Psi$ and $m_S$,
the heavy fermions $\Psi$ and the singlet scalar $S$ can be integrated out, while leaving  their footprints in the effective  interactions among the SM gauge bosons and Higgs boson. Then those effective interactions
eventually generate the nucleon and electron EDMs in the low energy limit through the loops involving the exchange of the SM gauge bosons and/or the Higgs boson.
In this process, one needs to take into account the renormalization group (RG) running, particularly those due to the QCD interactions, from the initial threshold scale   $m_\Psi\sim m_S$ down to the hadronic scale $\Lambda_{\rm QCD}$, as well as the intermediate threshold corrections from integrating out the massive SM particles.  

To simplify the calculation, we will ignore the RG running effects due to the QCD interactions over the scales from $m_\Psi$ to the SM Higgs boson mass
$m_H=125$ GeV. In this  approximation, the Wilsonian effective interactions at scales just below $m_H$
can be determined by the leading order Feynman diagrams  involving $\Psi, S$ and the SM Higgs boson.  We then take into account the subsequent RG running due to the QCD interactions  from $m_H$ to $\Lambda_{\rm QCD}$, while ignoring the threshold corrections due to the SM heavy quarks, to derive the low energy effective lagrangian at scales just above $\Lambda_{\rm QCD}$.

\subsection{Neutron EDM}

\begin{figure}[t]
\begin{center}
 \begin{tabular}{l}
  \includegraphics[scale=0.4]{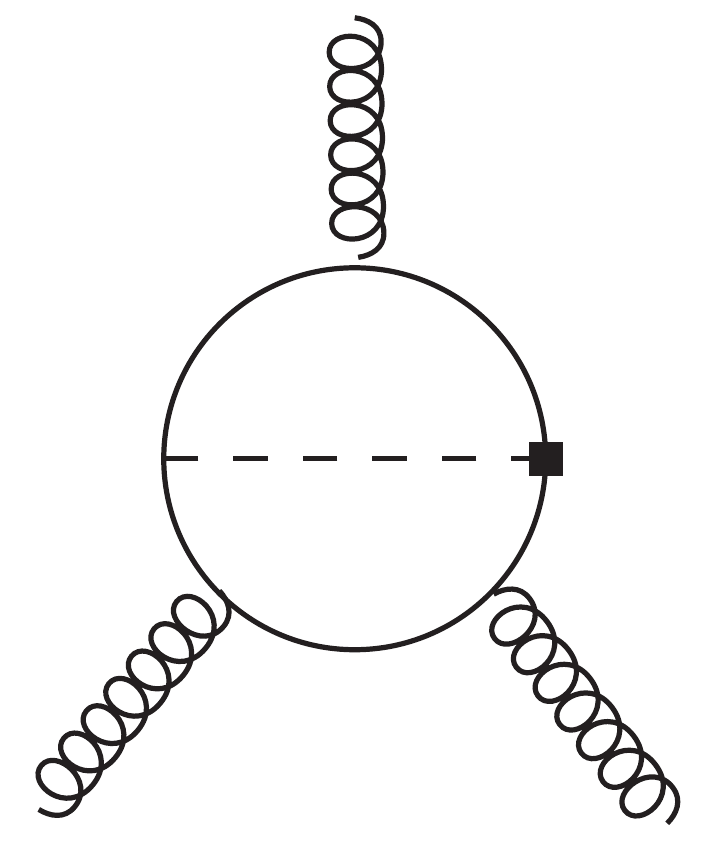}
   \end{tabular}
  \end{center}
  \caption{
The  Weinberg's three gluon interaction generated as a two-loop threshold correction. Here the small dark square represents
the $\gamma_5$-coupling of $S$ to the vector-like fermion $\Psi$. 
  }
\label{fig:weinberg_op}
\end{figure}

\begin{figure}[h]
\begin{center}
 \begin{tabular}{l}
  \includegraphics[scale=0.4]{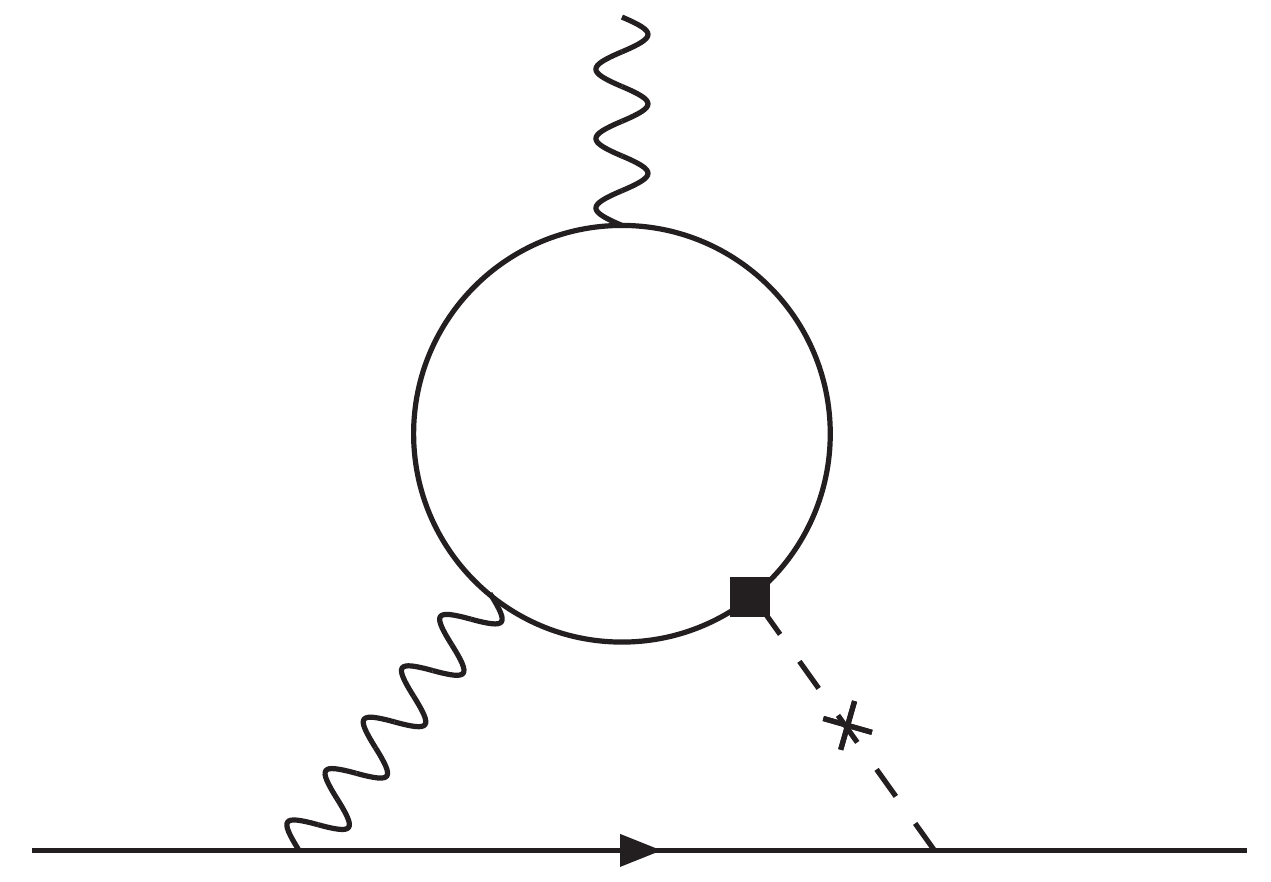}
  \hspace{1cm}
 \includegraphics[scale=0.4]{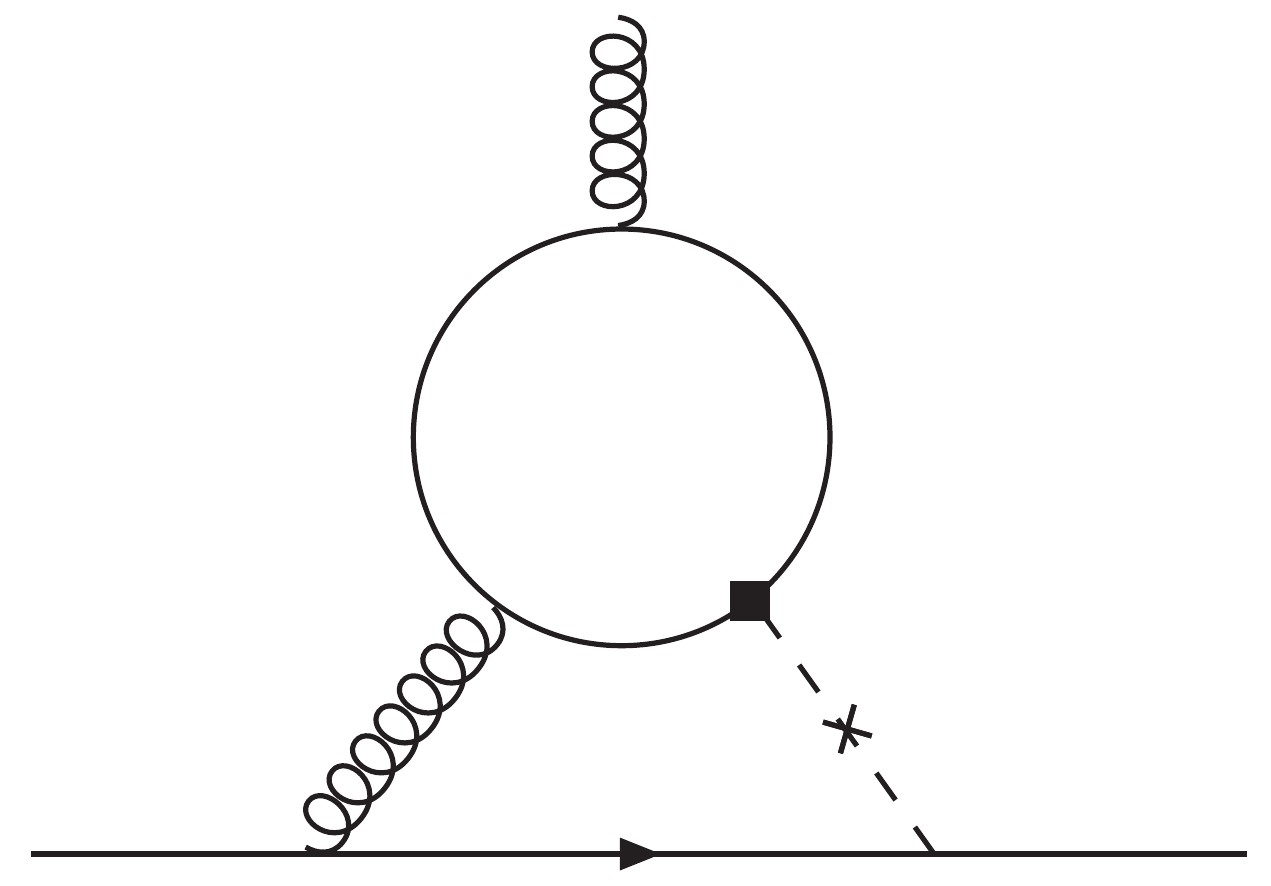}
   \end{tabular}
  \end{center}
  \caption{
The Barr-Zee diagrams for the EDM and chromo EDM (CEDM) of light fermions. The small cross denotes the $S-H$ mixing.
  }
\label{fig:Barr-Zee}
\end{figure}

The leading contribution to the neutron EDM turns out to come from the Weinberg's three gluon operator \cite{Weinberg:1989dx} generated by the diagram in Fig. \ref{fig:weinberg_op}.
In the presence of a mixing between the singlet scalar $S$ and the SM Higgs boson $H$, the EDM and chromo EDM (CEDM) of light quarks are induced by
the Barr-Zee diagrams \cite{Barr:1990vd} in Fig. \ref{fig:Barr-Zee}, which
may provide a potentially important  contribution to the neutron EDM.  

To be concrete, let us take a simple model having $N_F$ vector-like Dirac fermions $\Psi$ transforming under $SU(3)_c \times SU(2)_L \times U(1)_Y$ as
\dis{
\Psi = (3, 1)_{Y_\Psi},
}
where $Y_\Psi$ denotes the $U(1)_Y$ hypercharge of $\Psi$. As mentioned above, we take an approximation to ignore the RG running due to the QCD interactions between $m_\Psi$ and $m_H=125$ GeV.
Then at scales just below $m_H$,  the relevant Wilsonian effective interactions are determined to be
 \cite{Weinberg:1989dx,Abe:2013qla, Dicus:1989va, Jung:2013hka, Dekens:2014jka},
\bea
\label{eff_mH}
{\cal L_{\rm eff}}(m_H) &=&
-\frac{d_W(m_H)}{6} f_{abc}\epsilon^{\mu\nu\rho\sigma} G^a_{\rho \sigma} G^{b}_{\mu \lambda} G^{c\,\,\lambda}_{\nu} \nonumber \\
&&
 -\frac{i}{2} \sum_{q} \Big[ d_q(m_H) e \bar{q} \sigma^{\mu\nu} \gamma_5 q F_{\mu\nu} + \tilde{d}_q(m_H) g_3 \bar{q} \sigma^{\mu\nu} \gamma_5 T_3^a q G^{a\mu\nu} \Big]+...,\eea
with
\bea
d_q(m_H) &=& 4N_F \frac{e^2}{(4\pi)^4} \frac{m_q}{v}
\left( 6 Y_\Psi^2  \frac{y_S}{m_\Psi}  s_\alpha s_\xi c_\xi \right) 
\left[ Q_q +\left(  t_w^2 Q_q -\frac{T^3_{q_L}}{2 c_w^2} \right) \right]
\left[
g\left(\frac{m_\Psi^2}{m_{H}^2} \right) -
g\left(\frac{m_\Psi^2}{m_{S}^2} \right)  
\right], \label{dq_mpsi}
\nonumber\\
\tilde{d}_q(m_H) &=&  4 N_F \frac{g_3^2}{(4\pi)^4} \frac{m_q}{v}
\left( \frac{y_S}{m_\Psi}  s_\alpha s_\xi c_\xi \right)
\left[
g\left(\frac{m_\Psi^2}{m_{H}^2} \right) -
g\left(\frac{m_\Psi^2}{m_{S}^2} \right)  
\right], \label{tdq_mpsi}
\nonumber\\
d_W(m_H) &=& - N_F \frac{g_3^3}{(4\pi)^4} \frac{y_S^2}{m_\Psi^2} c_\alpha s_\alpha 
\left[ s_\xi^2 h\left(\frac{m_\Psi^2}{m_{H}^2} \right) + c_\xi^2 h\left(\frac{m_\Psi^2}{m_{S}^2}\right)
\right], \label{dW_mpsi}
\eea 
where $q=u,d,s$ stands for the light quark species, $s_\alpha=\sin\alpha$, $s_\xi=\sin\xi_{SH}$ for the $S-H$ mixing angle $\xi_{SH}$, $v=246$ GeV is the SM Higgs vacuum value, $c_w = \cos \theta_w$, $t_w = \tan \theta_w$ for the weak mixing angle $\theta_w$, and the loop functions $g$ and $h$ are given by\footnote{It is useful to note the aymptotic behavior of the loop functions: $h(z\ll1) \simeq z \ln (1/z)$, $h(z\gg 1) \simeq 1/4$, and $g(z \gg1 ) \simeq 1 + (\ln z)/2$.}
\bea
g(z) &\equiv&  \frac{z}{2} \int^1_0 dx \frac{1}{x(1-x) - z} \ln \frac{ x(1-x)}{z}, \nonumber \\
h(z) &\equiv& z^2 \int^1_0 dx \int^1_0 dy \frac{x^3y^3 (1-x)}{[z x(1-xy) + (1-x)(1-y)]^2}.
\label{g_h}
\eea
Let us recall that the parameter ratio $m_\Psi/y_S$ has a specific connection with the diphoton cross section
$\sigma(pp\rightarrow \gamma\gamma)$, which is given by (\ref{m/y}). This allows us to estimate  the expected size of the EDMs in terms of a few model parameters such as $\alpha$ and $\xi_{SH}$.

In order to estimate the resulting neutron EDM, we should bring the effective interactions (\ref{eff_mH}) down to the QCD scale through the RG evolution. 
For this, it is convenient to redefine the coefficients as
\bea
C_1(\mu) = \frac{d_q(\mu)}{m_q Q_q}, 
\quad 
{C}_2(\mu) = \frac{\tilde{d}_q(\mu)}{m_q},
\quad 
C_3(\mu) = \frac{d_W(\mu)}{g_3},
\eea
which are satisfying the RG equation \cite{Degrassi:2005zd, Hisano:2012cc}:
\bea
\mu \frac{\partial {\bf C}}{\partial \mu} = \frac{g_3^2}{16\pi^2} \gamma \, {\bf C},
\eea
with the anomalous dimension matrix
\bea
\gamma \equiv
\left(
\begin{array}{ccc}
\gamma_e &\gamma_{eq}  & 0
\\
0 & \gamma_q & \gamma_{Gq}
\\
0 &0 & \gamma_G
\end{array}
\right)
= 
\left( 
\begin{array}{ccc}
8 C_F & 8 C_F & 0 \\
0 & 16C_F -4 N_c &  2 N_c \\
0 & 0 & N_c + 2 n_f +\beta_0
\end{array}
\right),
\eea
where ${\bf C} = ( C_1, {C}_2,C_3)^T$,
 $N_c=3$ is the number of color, $C_F = 4/3$ is a quadratic Casimir, $n_f$ is the number of active light quarks, and $\beta_0 = (33-2n_f)/3$ is the one-loop beta function coefficient.
Solving  this RG equations, one finds \cite{Degrassi:2005zd}
\bea
C_1 (\mu) &=&
\eta^{\kappa_e} C_1(m_H) + \frac{\gamma_{qe}}{\gamma_e - \gamma_q} (\eta^{\kappa_e} - \eta^{\kappa_q} )C_2(m_H) 
\nonumber \\
&+& 
\left[
\frac{\gamma_{Gq}\gamma_{qe} \eta^{\kappa_{e}} }{(\gamma_q - \gamma_e)(\gamma_G -\gamma_e)} 
+ 
\frac{\gamma_{Gq}\gamma_{qe} \eta^{\kappa_{q}} }{(\gamma_e - \gamma_q)(\gamma_G -\gamma_q)}  
+ 
\frac{\gamma_{Gq}\gamma_{qe} \eta^{\kappa_{G}} }{(\gamma_e - \gamma_G)(\gamma_q -\gamma_G)} 
\right] C_3(m_H), \nonumber
\\
C_2(\mu) &=& 
\eta^{\kappa_q}C_2(m_H) 
+ \frac{\gamma_{Gq}}{\gamma_q - \gamma_G} 
\left[
\eta^{\kappa_q} - \eta^{\kappa_G} \right] C_3(m_H), \nonumber 
\\
C_3(\mu) &=& \eta^{\kappa_G} C_3(m_H), \label{C3}
\eea
where $\eta \equiv g_3^2(m_H) / g_3^2(\mu)$ and $\kappa_x = \gamma_x / (2\beta_0)$. 
The analytic expressions for $C_i (\mu \sim \Lambda_{\rm QCD} )$ in terms of $C_i (m_H)$ are complicated except $C_3$, however fortunately it turns out that the dominant contribution to the neutron EDM comes from $C_3(\mu\sim \Lambda_{\rm QCD})$.
From (\ref{C3}), we obtain
\dis{ \label{dW_mu}
d_W(\mu) = \left(\frac{g_3(m_c)}{g_3(\mu)}\right)\left(\frac{g_3(m_b)}{g_3(m_c)}\right)^\frac{33}{25}
\left(\frac{g_3(m_H)}{g_3(m_b)}\right)^\frac{39}{23} d_W(m_H).
}
It can be  shown numerically that  $d_q(\mu)$ and $\tilde d_q(\mu)$ also get a similar amount of suppression by the RG evolution compared to the high scale values at $m_H$.

Now one can relate the Wilsonian coefficients $d_W(\mu), d_q(\mu)$ and $\tilde d_q(\mu)$ at $\mu\sim \Lambda_{\rm QCD}$
to the neutron EDM:
\bea
-\frac{i}{2}d_n \bar{n}\sigma^{\mu\nu}\gamma_5 n F_{\mu\nu},\eea
which is the most ambiguous  step.
For this, one can take two approaches, the Naive Dimensional Analysis (NDA)  \cite{NDA} or the QCD sum rule \cite{Pospelov:2000bw, Hisano:2012sc, Hisano:2015rna}, essentially yielding similar results.
As  for the neutron EDM estimated by the NDA, one finds \bea
d_n/e =  {\cal O}(d_q(\mu)) + {\cal O}(\tilde d_q(\mu)/\sqrt{6}) +  {\cal O}(f_\pi d_W(\mu)),\eea
where the corresponding scale $\mu$ is chosen to be the one with
$g_3(\mu)\simeq 4\pi/\sqrt{6}$ \cite{Weinberg:1989dx}.
On the other hand, applying the QCD sum rule for the neutron EDM $d^q_n$ from the (C)EDM of light quarks, one finds a more concrete 
result\footnote{We are using ``the modified QCD sum rule" obtained by assuming the Peccei-Quinn mechanism to dynamically cancel the QCD vacuum angle.} \cite{Hisano:2015rna}:
\bea \label{dn_dd}
d^q_n/e \simeq -0.2 d_u(\mu) + 0.78 d_d(\mu) + 0.29 \tilde{d}_u(\mu) + 0.59 \tilde{d}_d(\mu).
\eea
for $\mu \simeq 1$ GeV. As for the neutron EDM $d^W_n$ from the Weinberg's three gluon operator in the QCD sum rule approach, one similarly finds \cite{Demir:2002gg}
\bea
\label{dn_dW}
|d^W_n/e| = \left(1.0^{+1.0}_{-0.5} \right)\times 20\, {\rm MeV} \times |d_W(\mu)|\eea
for $\mu \simeq 1$ GeV.
We can now make a comparison between  the neutron EDM $d_n^W$ originating from  $d_W(\mu)$ and the other part
$d_n^q$ originating from $d_q(\mu)$ and $\tilde d_q(\mu)$. Within the QCD sum rule approach, we find numerically
\dis{
{d_n^q}/{d_n^W} 
\,\simeq \, 3 \, \sin \xi_{SH} + 0.07.
}
This implies that the neutron EDM is dominated by the contribution from the Weinberg's three gluon operator for  the $S-H$ mixing angle $\xi_{SH}\lesssim 0.1$, which might be required to be consistent with the Higgs precision data \cite{Falkowski:2015swt,Cheung:2015cug}\footnote{If one uses the NDA rule or the chiral perturbation theory \cite{Fuyuto:2012yf}, the resulting neutron EDM induced by the (C)EDM of the strange quark can be comparable to the contribution from the Weinberg's three gluon operator for 
the $S-H$ mixing angle $\xi_{SH} \sim 0.1$.}.

\begin{figure}[t]
\centering
\includegraphics[scale=0.5]{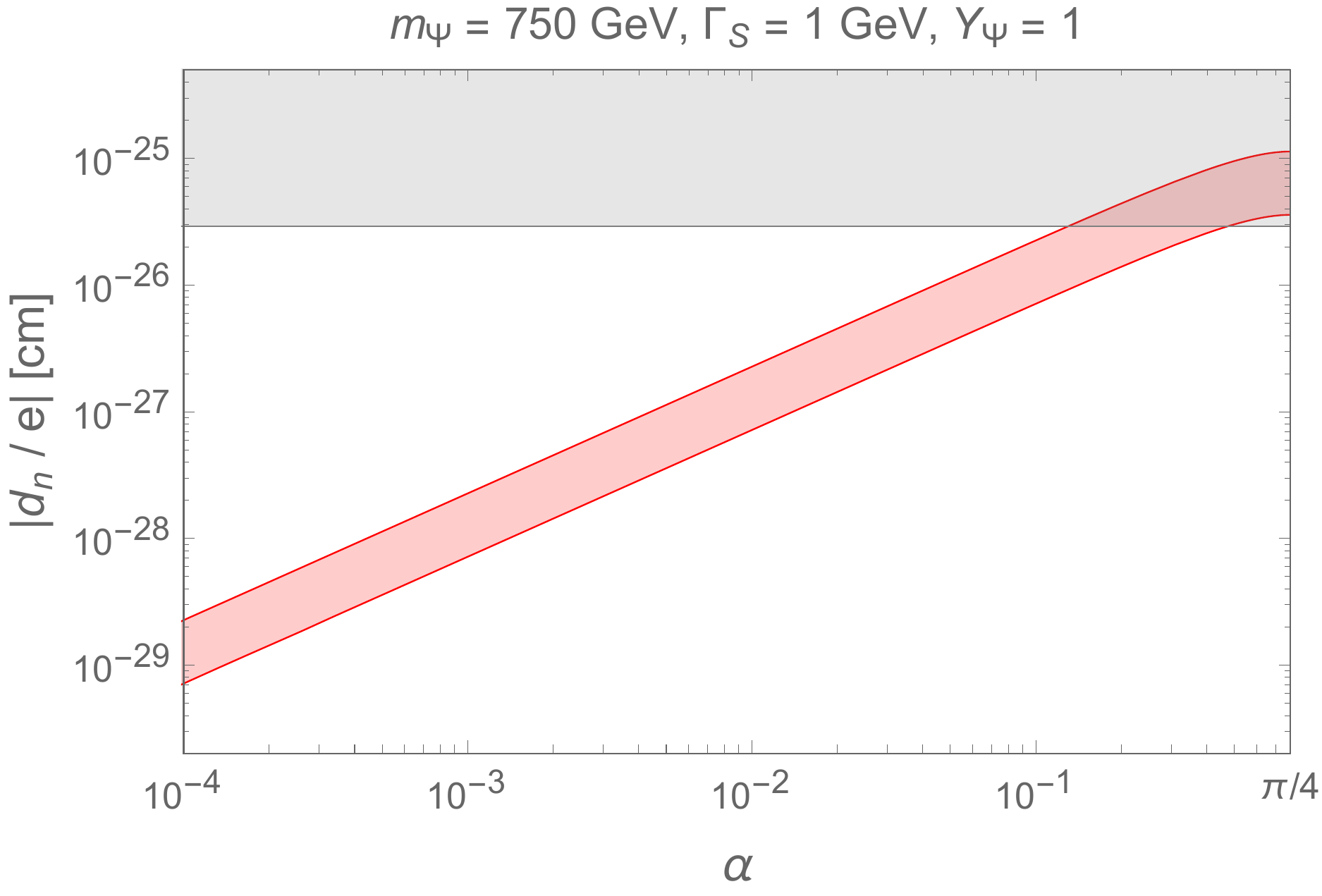}
\caption{The neutron electric dipole moment as a function of the CP violating angle $\alpha$ for the model parameters to give $\sigma_{\rm signal} = 1 - 10$ fb. For this plot, we choose the total decay width of $S$ as $\Gamma_S = 1$ GeV, the number of Dirac fermions $\Psi$ as $N_F=1$, the mass and $U(1)_Y$ hyperchage of $\Psi$ as  $m_\Psi = 750$ GeV and $Y_\Psi =1$. }
\label{fig:dnEst}
\end{figure}

With the above observation,
plugging (\ref{m/y}), (\ref{dW_mpsi}) and (\ref{dW_mu}) into (\ref{dn_dW}), we obtain the following expression for the expected neturon
EDM over the 750 GeV signal region:
\dis{
d_n /e \simeq 3 \times 10^{-25} \, {\rm cm} &\times \frac{ c_\alpha s_\alpha}{N_F Y_\Psi^2} \sqrt{\left( \frac{\Gamma_S}{1\, \rm GeV} \right) \left( \frac{\sigma_{\rm signal}}{8 \, \rm fb} \right)} \times  R_n,}
where
$$
R_n =
\left(\frac{h(4\tau_\Psi)}{h(1)}\right) \left[ \frac{c_{0.1}^2 (A_{1/2}(1/4)/2)^2 + s_{0.1}^2 (4f(1/4))^2}{c_\alpha^2 (A_{1/2}(\tau_\Psi)/2)^2 + s_\alpha^2 (f(\tau_\Psi)/\tau_\Psi)^2}\right]^{1/2}=\, {\cal O}(1).
$$
Here $R_n$ represents the dependence on the loop functions $A_{1/2}, f$ and $h$ defined in 
(\ref{A-f}) and (\ref{g_h}), which is normalized to the value at $\tau_\Psi = m_S^2/4m_\Psi^2 = 1/4$ and $\alpha = 0.1$. 
With this result, one can easily see that the neutron EDM from the 750 GeV sector saturates the current experimental upper bound 
$\sim 3 \times 10^{-26} \, {\rm e\cdot cm}$ 
\cite{Baker:2006ts} for the parameter region with $\sin2\alpha/N_FY_\Psi^2  \sim 0.1$.
In Fig. \ref{fig:dnEst}, we depict the resulting neutron EDM as a function of CP violating angle $\alpha$ for the model parameters which give the diphoton cross section $\sigma(pp\rightarrow \gamma\gamma))=1\sim 10$ fb. 

\subsection{Electron EDM}

\begin{figure}
\centering
\includegraphics[scale=0.5]{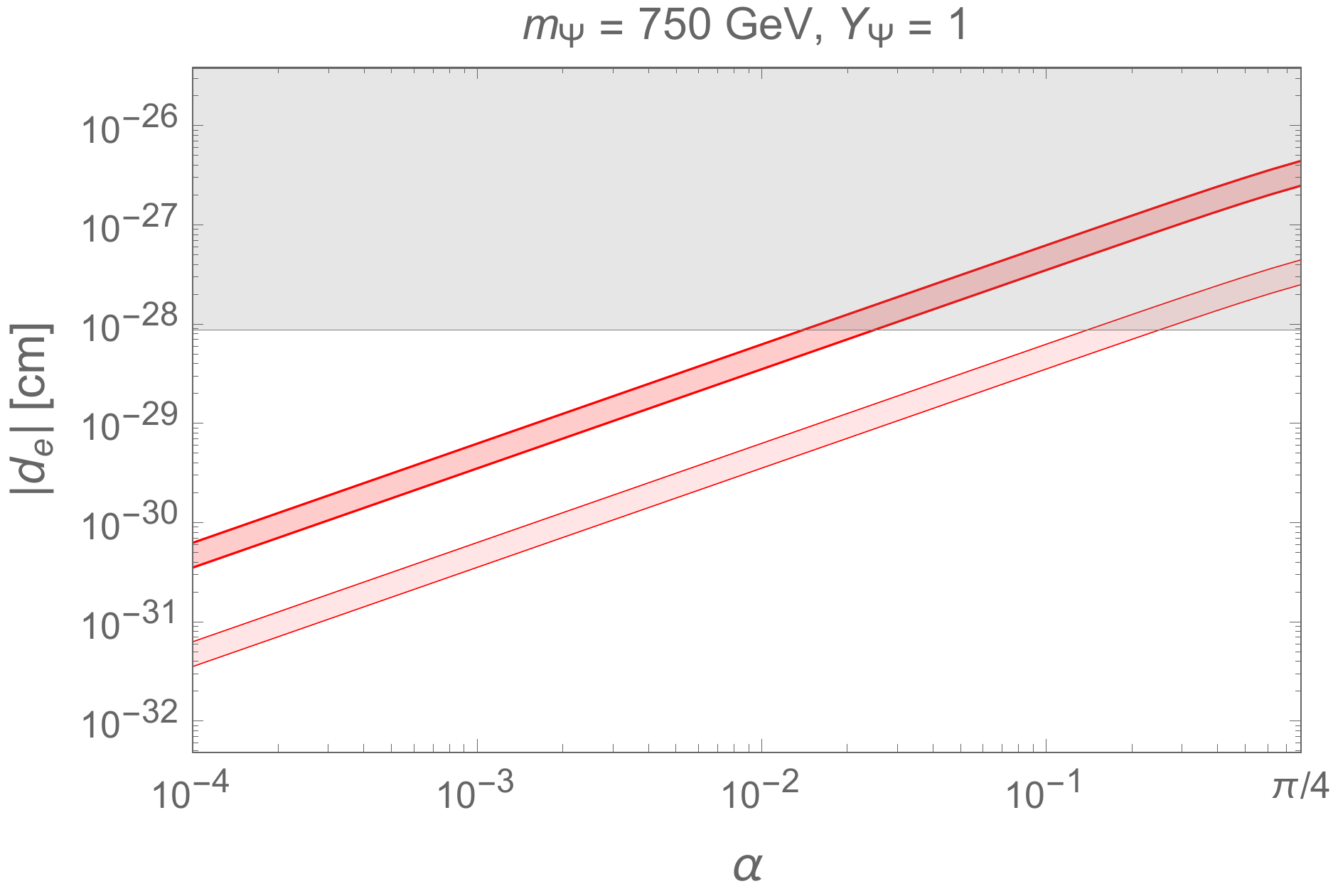}
\caption{ The electron electric dipole moment  for the minimal model with $\Psi = (3,1)_{Y_\Psi}$,
$m_\Psi = 750$ GeV, $\xi_{SH} = (10^{-1}, 10^{-2})$, $Y_\Psi =1$, and $\sigma_{\rm signal} = 1-10$ fb. }
\label{fig:electron EDM}
\end{figure}

In the presence of the $S-H$ mixing, a sizable electron EDM  can arise from the Barr-Zee diagram in Fig. (\ref{fig:Barr-Zee}).
In case of the model with $N_F$ flavors of $\Psi = (3, 1)_{Y_\Psi}$, we obtain the electron EDM
\dis{
 -\frac{ie}{2}  d_e(\mu)  \bar{e} \sigma^{\mu\nu} \gamma_5 e F_{\mu\nu},
 }
 with the coefficient \cite{Jung:2013hka, Dekens:2014jka}
\bea
d_e &=& -24 N_F \frac{e^2}{(4\pi)^4} \frac{m_e}{v} 
\Big( Y_{\Psi}^2\frac{y_S}{m_{\Psi}} s_\alpha s_\xi c_\xi \Big)
\Big( 1 + t_w^2 - \frac{1}{4 c_w^2} \Big)
\left[
g\left(\frac{m_\Psi^2}{m_{H}^2}\right)
-g\left(\frac{m_\Psi^2}{m_{S}^2}\right)
\right],
\label{de_BZ}
\eea
where the loop function $g(z)$ is given in (\ref{g_h}) and the other parameters are defined as same as
in (\ref{dW_mpsi}).
Applying the relation (\ref{m/y}) for the above result, we find
\dis{
\label{de_mixing}
d_e =-( 6.2 &\times 10^{-26} \,{\rm cm} )\times s_\alpha s_\xi c_\xi Y_\Psi 
\left( \frac{\Gamma_S}{1\, \rm GeV}\right)^{1/4} \left( \frac{\sigma_{\rm signal}}{8 \, \rm fb} \right)^{1/4}\times R_e,}
where
$$R_e = \left( \frac{g(m_S^2/4\tau_\Psi m_H^2) - g (1/4\tau_\Psi)}{g(m_S^2/m_H^2) - g(1)}\right)
 \left( \frac{c_{0.1}^2 (A_{1/2}(1/4)/2)^2 + s_{0.1}^2 (4f(1/4))^2}{c_\alpha^2 (A_{1/2}(\tau_\Psi)/2)^2 + s_\alpha^2 (f(\tau_\Psi)/\tau_\Psi)^2} \right)^{1/2}=\,{\cal O}(1)
 $$
 for $\tau_\Psi = m_S^2/4m_\Psi^2$.
The above result shows the electron EDM associated with the $S-H$ mixing can saturate the current experimental upper limit $8.7 \times 10^{-29}$ cm \cite{Baron:2013eja}
when $\sin\alpha\sin\xi_{SH} ={\cal O}(10^{-3})$.
In  Fig. (\ref{fig:electron EDM}), we depict 
the electron EDM over the 750 GeV signal region for the two different values of the $S-H$ mixing angle: $\xi_{SH}=10^{-1}$ and $10^{-2}$. 

If the vector-like fermions $\Psi$ carry a nonzero $SU(2)_L$ charge, there can be a nonzero electron EDM even in the limit $\xi_{SH}=0$.  
For instance, in the model with $N_F$ flavors of $\Psi = (3, 2)_{Y_\Psi}$,  a CP-odd three $W$-boson operator
of the form
$$
\frac{\tilde d_W}{3}\epsilon_{ijk}W^{i}_{\mu\rho} W_\nu^{j \,\, \rho}\widetilde{W}^{k\mu\nu}
$$
can be generated by the loops of $\Psi$.
Following \cite{Marciano:1986eh,Boudjema:1990dv}\footnote{The authors in \cite{Boudjema:1990dv} noticed  that the result is scheme-dependent. This means that the precise result depends on the dependence of $\tilde d_W$ on the external $W$-boson momenta. 
Here we simply use the result from the dimensional regularization for the purpose of estimation of the electron EDM.}, we find the resulting electron EDM is given by
\dis{
\label{de_3W}
\Delta d_e & \simeq  - \frac{N_F}{4} \frac{g_2^4}{(16 \pi^2)^3} m_e \frac{y_S^2}{m_\Psi^2} c_\alpha s_\alpha \left[ s_\xi^2 h\left(\frac{m_{H}^2}{m_\Psi^2}\right) + c_\xi^2 h\left(\frac{m_S^2}{m_\Psi^2}\right) \right] \\
&\simeq - (6.5 \times 10^{-31} \, {\rm cm} )\times \frac{c_\alpha s_\alpha}{N_F Q_\Psi^2} \sqrt{\left( \frac{\Gamma_S}{1\, \rm GeV} \right) \left( \frac{\sigma_{\rm signal}}{8 \, \rm fb} \right)} \times \tilde R_e,}
where
$$
\tilde R_e = \left(\frac{c_\xi^2 h(4\tau_\Psi)+s_\xi^2 h(m_H^2/m_\Psi^2)}{c_{0.1}^2 h(1)+s_{0.1}^2 h(m_H^2/m_S^2)}\right) \left( \frac{c_{0.1}^2 (A_{1/2}(1/4)/2)^2 + s_{0.1}^2 (4f(1/4))^2}{c_\alpha^2 (A_{1/2}(\tau_\Psi)/2)^2 + s_\alpha^2 (f(\tau_\Psi)/\tau_\Psi)^2}\right)^{1/2}=\, {\cal O}(1).
$$
For $\sin 2\alpha \lesssim 0.1$, which might be required to satisfy the bound on the neutron EDM, the resulting electron EDM is about three orders of magnitude smaller than the current bound, therefore too small to be observable in a foreseeable future.

\section{Composite pseudo-Nambu-Goldstone resonance} \label{sec:composite}

In the previous section, we discussed the neutron and electron EDM in models
where the 750 GeV resonance is identified as an elementary spin zero field  (at least at scales around TeV) which couples to vector-like fermions to generate the effective couplings to
explain the diphoton excess $\sigma(pp\rightarrow \gamma\gamma)\sim 5\,\, {\rm fb}$.  
On the other hand, it has been pointed out that in most cases this scheme confronts with a strong coupling regime at scales not far above the TeV scale
\cite{Gu:2015lxj, Son:2015vfl}. In regard to this, an interesting possibility is that $S$ corresponds to a composite pseudo-Nambu-Goldstone (PNG) boson of the spontaneously broken chiral symmetry of a new QCD-like hypercolor dynamics which confines at $\Lambda_{\rm HC}={\cal O}(1)$ TeV~\cite{Harigaya:2015ezk,Nakai:2015ptz,Redi:2016kip}. As is well known, such models involve a unique source of CP violation, the hypercolor vacuum angle
$\theta_{\rm HC}$, which can yield a nonzero neutron or electron EDM in the low energy limit \cite{Harigaya:2015ezk}.

To proceed, we consider a specific example, the model discussed  in \cite{Nakai:2015ptz}, involving a hypercolor gauge group $SU(N)_{\rm HC}$ with charged Dirac fermions $(\psi, \chi)$ which transform under $SU(N)_{\rm HC}\times SU(3)_c\times SU(2)_L\times U(1)_Y$ as
\bea
\psi
=(N, 3, 1)_{Y_\psi}, \quad  \chi=(N,1, 1)_{Y_\chi},\eea where
 $Y_{\psi,\chi}$ denote the $U(1)_Y$ hypercharge. 
 At scales above $\Lambda_{HC}$, the lagrangian of the hypercolor color sector is given by 
\bea
{\cal L}_{\rm HC}&=&-\frac{1}{4g_{\rm HC}^2}H^{a\mu\nu}H^a_{\mu\nu}-\frac{\theta_{\rm HC}}{32\pi^2}H^{a\mu\nu}\widetilde{H}^a_{\mu\nu}\nonumber \\
&+&\bar\psi i \slashed{D} \psi +\bar\chi i \slashed{D}\chi-\bar\psi m_\psi\psi -\bar\chi m_\chi\chi,
\eea
where $H^{a\mu\nu}$ denotes the $SU(N)_{\rm HC}$ gauge field  strength,  $\widetilde H^{a\mu\nu}$ is its dual, and
the fermion masses $m_{\psi,\chi}$ are chosen to be real and $\gamma_5$-free.
For a discussion of the low energy consequence of the CP-violating vacuum angle $\theta_{\rm HC}$,  it is convenient to make
a chiral rotation of fermion fields to rotate away $\theta_{\rm HC}$ into the phase of the fermion mass matrix, which results in
\bea
M= M_\theta \equiv \mbox{diag}\left(m_\psi e^{ix_\psi\theta_{\rm HC}},
m_\psi e^{ix_\psi\theta_{\rm HC}},m_\psi e^{ix_\psi\theta_{\rm HC}},m_\chi e^{ix_\chi\theta_{\rm HC}}\right),\eea
where
$$3x_\psi+x_\chi=1.
$$

For $m_{\psi,\chi}\ll \Lambda_{\rm HC}$,  the model is invariant under an approximate chiral symmetry $SU(4)_L\times SU(4)_R$ which is spontaneously broken down to the diagonal $SU(4)_V$ by the fermion bilinear condensates:
\bea
|\langle \bar\psi_L \psi_R\rangle |\,\simeq\,
 |\langle \bar\chi_L \chi_R\rangle |\, \simeq\,  \frac{N}{16\pi^2}\Lambda_{\rm HC}^3.\eea 
The corresponding pseudo-Nambu-Goldstone (PNG) boson can be described by an $SU(4)$-valued field $U=\exp(2 i\Phi/f)$ whose low energy dynamics is governed by 
\bea
{\cal L}_{\rm eff}=\frac{1}{4}f^2\mbox{tr}\left(D_\mu U D^\mu U^\dagger\right) +  \mu^3\mbox{tr}\left(M_\theta U^\dagger + \mbox{h.c}\right) +{\cal L}_{\rm WZW}+{\cal L}_{\rm CPV}...,  \eea
where the naive dimensional analysis suggests 
\bea
f^2\,\simeq \frac{N}{16\pi^2}\Lambda_H^2, \quad \mu^3 \simeq \frac{N}{16\pi^2}\Lambda_H^3, \eea
and ${\cal L}_{\rm WZW}$ and ${\cal L}_{\rm CPV}$ denote the Wess-Zumino-Witten term and the additional CP-violating terms, respectively.  
For a discussion of CP violation due to $\theta_{\rm HC}\neq 0$, it is convenient to choose the fermion mass matrix $M_\theta$ as
\bea
-\frac{i}{2}\left(M_\theta-M_\theta^\dagger\right) = m_\theta {\bf 1}_{4\times 4},\eea
for which 
 the PNG boson
has a vanishing vacuum expectation value.
Then the CP violation due to $\theta_{\rm HC}\neq 0$ is parametrized simply by  
\bea
m_\theta \equiv m_\psi \sin \left(x_\psi{\theta}_{\rm HC}\right) =m_\chi \sin\left(x_\chi {\theta}_{\rm HC}\right) \quad \left(3x_\psi+x_\chi=1\right), \eea
which manifestly shows that CP is restored if $\theta_{\rm HC}$ or any of $m_{\psi,\chi}$  is vanishing. In the limit
$|\theta_{\rm HC}|\ll 1$, this order parameter for CP violation has a simple expression:
\bea
m_\theta \,\simeq \, \frac{\theta_{\rm HC}}{\mbox{tr}\left(M^{-1}_{\bar\theta_H=0}\right)} =\frac{m_\psi m_\chi}{3m_\chi+m_\psi}\theta_{\rm HC}.
\eea
The PNG bosons of $SU(4)_L\times SU(4)_R/SU(4)_V$ includes a unique SM-singlet component $S$ which can be identified as the 750 GeV resonance:
\bea
\Phi = \frac{1}{2\sqrt{6}}{\rm diag}(S,S,S, -3S) + \cdots,
\eea
where $U=\exp(2 i\Phi/f)$, and
the ellipsis denotes the $SU(3)_c$ octet and triplet PNG bosons which are heavier than $S$. 
Then the Wess-Zumino-Witten term gives rise to the following effective couplings between $S$ and the SM gauge bosons,
which would explain the diphoton excess:
\bea
{\cal L}_{\rm WZW} = -\frac{N}{16\pi^2}\frac{S}{f}\left( \frac{1}{2\sqrt{6}} g_3^2 G^{a\mu\nu}\widetilde G^a_{\mu\nu} + \frac{\sqrt{6}}{2} {g_1}^2 \left(Y_\psi^2-Y_\chi^2\right)  B^{\mu\nu}\widetilde B_{\mu\nu}\right)+ \cdots.
\eea
With $m_\theta\neq 0$, according to the NDA, the underlying hypercolor dynamics generates the following CP violating effective interactions renormalized at $\Lambda_{\rm HC}$: 
\bea
{\cal L}_{\rm CPV} &=&\frac{N}{16\pi^2}\frac{m_\theta}{\Lambda_H}\frac{S}{f}
\left( c_Gg_3^2G^{a\mu\nu} G^a_{\mu\nu} + c_B{g^\prime}^2\left(Y_\psi^2-Y_\chi^2\right) B^{\mu\nu} B_{\mu\nu}\right)\nonumber \\
&+& \frac{N}{16\pi^2}\frac{m_\theta}{\Lambda_H} \frac{\kappa_G}{\Lambda_H^2}\frac{g_3^3f_{abc}}{3}G^{a}_{\mu\rho}G^{b\,\,\rho}_{\nu}\widetilde G^{c \, \mu\nu}+ \cdots, \eea
where $c_G, c_B$ and $\kappa_G$ are all of order unity.

\begin{figure}
\centering
\includegraphics[scale=0.5]{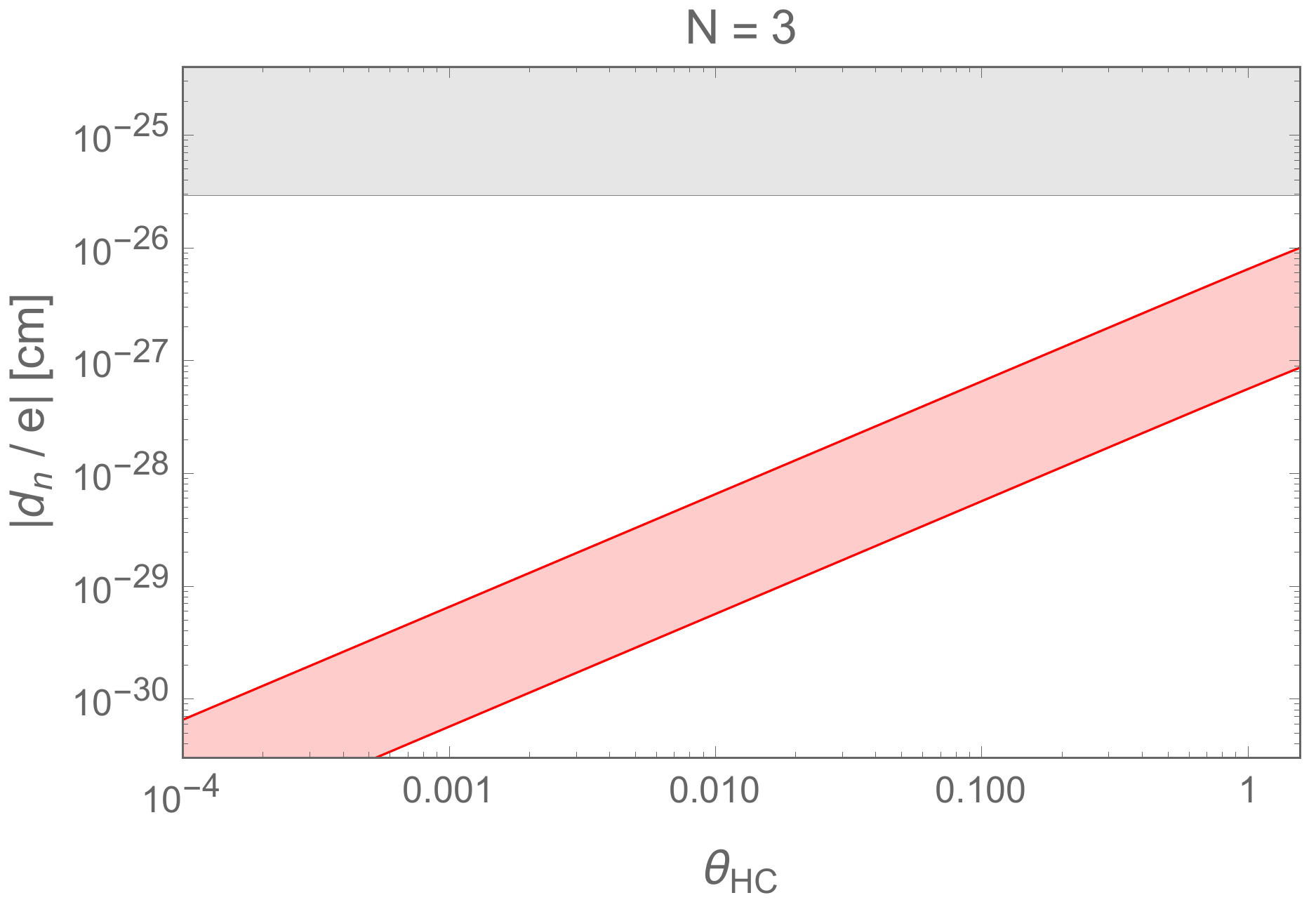}
\caption{{The expected neutron electric dipole moment as a function of the hypercolor  vacuum angle $\theta_{\rm HC}$ in models for  a composite PNG 750 GeV resonance. For this plot, we choose $N=3$.}}  
\label{fig:Composite_nEDM}
\end{figure}

It is now straightforward to use our previous results to find the nucleon and electron EDM induced by the above effective interactions. 
By matching the coefficients of the relevant interactions with the simple model presented in section \ref{sec:model}, we find the following correspondence:
\dis{
\label{corres}
\frac{y_S}{m_\Psi}\sim \frac{N}{f} , \quad \sin 2\alpha \sim \frac{m_S^2}{\Lambda_{\rm HC}^2} \sin {\theta}_{\rm HC},
}
where we have used  the relation
\bea
m_S^2 = (750\,\, {\rm GeV})^2 \simeq \frac{(m_\psi + 3m_\chi) \mu^3}{f^2} \simeq (m_\psi+ 3m_\chi) \Lambda_{\rm HC}.
\eea
Since the ratio $m_\Psi/ y_S$ should be around 100 GeV  to explain the 750 GeV diphoton excess,  it turns out that
$f\sim N\times 100$ GeV and $\Lambda_{\rm HC}\simeq 4\pi f/\sqrt{N}\sim \sqrt{N}$ TeV, implying that roughly $\sin 2\alpha$ is one or two orders of magnitude smaller than $\theta_{\rm HC}$.
In Fig.~\ref{fig:Composite_nEDM}, we depict the neutron EDM  in the minimal model for a composite  PNG 750 GeV resonance for  the parameter region to give $\sigma(pp\rightarrow \gamma\gamma)=1\sim 10$ fb.

The minimal model of \cite{Nakai:2015ptz} can be generalized or modified to include  a hypercolored fermion carrying a nonzero $SU(2)_L$
charge \cite{Harigaya:2015ezk}, e.g. $\chi$ can transform as $(N,1,2)_{Y_\chi}$ under $SU(N)_H\times SU(3)_c\times SU(2)_L\times U(1)_Y$. Then the hypercolor dynamic with nonzero $\theta_{\rm HC}$ can generate the following CP-odd three $W$-boson operator:
\bea
\Delta {\cal L}_{\rm CPV} = \frac{N}{16\pi^2}\frac{m_\theta}{\Lambda_H}\frac{\kappa_W}{\Lambda_H^2}\frac{g_2^3 \epsilon_{ijk}}{3}W^{i}_{\mu\rho}W^{j\,\,\rho}_{\nu}\widetilde{W}^{k\,\mu\nu},
\eea
where $\kappa_W={\cal O}(1)$ according to the NDA rule.
Applying our previous result (\ref{de_3W}) under the relation (\ref{corres}), we find the electron EDM resulting from the above three $W$-boson operator is too small to be observable  even when $\theta_{\rm HC}={\cal O}(1)$.

Finally let us note that a composite PNG boson $S$ can have a mixing with the SM Higgs boson if the underlying hypercolor model
includes a higher dimensional operator of the form:
\bea
\frac{1}{\Lambda_\psi}|H|^2\bar \psi_L\psi_R + \frac{1}{\Lambda_\chi}|H|^2\bar\chi_L\chi_R +{\rm h.c},
\eea 
where $\Lambda_{\psi,\chi}$ are complex in general. For instance, this form of dim$-5$ operators can be generated by an exchange
of heavy scalar field  $\sigma$ which has the couplings 
\bea
{\cal L}_\sigma = -\frac{1}{2}m_\sigma^2 \sigma^2 + A_\sigma \sigma |H|^2 + \left(\lambda_\psi \sigma\bar \psi_L\psi_R +\lambda_\chi\sigma \bar\chi_L\chi_R+{\rm h.c.}\right) + ...,
\eea
yielding
$$
\frac{1}{\Lambda_\psi}=\frac{A_\sigma\lambda_\psi}{m_\sigma^2},\quad \frac{1}{\Lambda_\chi}=\frac{A_\sigma\lambda_\chi}{m_\sigma^2}.$$
The resulting $S-H$ mixing angle is estimated as 
\bea\xi_{SH} \, \sim \, \frac{v\Lambda_{\rm HC}^2}{4\pi m_S^2}\, \frac{{\rm Im}(\Lambda_{\psi,\chi})}{|\Lambda_{\psi,\chi}|^2},
\eea
where $v=246$ GeV is the vacuum value of the SM Higgs doublet $H$. 
One can apply this the mixing angle for our previous result (\ref{de_mixing}) to estimate the resulting electron EDM. 
Note that here  $S$ is an approximately pseudoscalar boson, and therefore  $\xi_{SH}$ corresponds to a CP-violating mixing angle, while $\sin\alpha$ is CP-conserving and of order unity.
One then finds the current bound on the electron EDM implies
\bea
\frac{\Lambda_{\rm HC}}{|\Lambda_{\psi,\chi}|}\frac{{\rm Im}(\Lambda_{\psi,\chi})}{|\Lambda_{\psi,\chi}|} \,\lesssim\, {\cal O}(10^{-2}).\eea

\section{Conclusion} \label{sec:conc}

The recently announced diphoton excess at 750 GeV in the Run II ATLAS and CMS data may turn out to be the first discovery of new physics beyond the Standard Model at collider experiments. In this paper, we examined the implication of the 750 GeV diphoton excess for the EDM of neutron and electron in models in which the diphoton excess is due to a spin zero resonance $S$ which couples to photons and gluons through the loops of massive vector-like fermions.
We found that a neutron EDM comparable to the current experimental bound can be obtained if the CP violating order parameter $\sin 2\alpha$ in the underlying new physics  is of ${\cal O}(10^{-1})$.
An electron EDM near the present bound can be obtained also when $\sin\xi_{SH}\times \sin\alpha ={\cal O}(10^{-3})$, where $\xi_{SH}$ is the mixing angle between $S$ and the SM Higgs boson.
For the case that $S$ corresponds to a pseudo-Nambu-Goldstone boson of a QCD-like hypercolor dynamics, one can use the correspondence $\sin 2\alpha \sim m_S^2\sin \theta_{\rm HC}/\Lambda^2_{\rm HC}$ to estimate the resulting EDMs, where
$\Lambda_{\rm HC}$ is the scale of spontaneous chiral symmetry breaking by the hypercolor dynamics and $\theta_{\rm HC}$ is the hypercolor vacuum angle. In view of that a nucleon or electron EDM near the current bound can be obtained over a natural parameter region of the model,  future precision measurements of the nucleon or electron EDM are highly motivated.

\section{Acknowledgment}
This work was supported by IBS under the project code, IBS-R018-D1.

\end{document}